\documentstyle[twocolumn,aps]{revtex}

\renewcommand{\abstractname}{ }
\makeatletter
\long\def\abstract#1{\def\@abstract{#1}}
\def\@abstract{}
\let\@oldmaketitle\@maketitle
\def\@maketitle{
 \@oldmaketitle
 \centerline{\bfseries\abstractname}
 \begin{quotation}\@abstract\end{quotation}
 \vskip-5mm}
\makeatother
\begin{document}
\draft
\preprint{}
\title{
Electronic Structure of B-2$p\sigma$ and $p\pi$ States in MgB$_2$, AlB$_2$ 
and ZrB$_2$ Single Crystals
}
\author{Jin Nakamura, Shin-ya Nasubida, Eiki Kabasawa, Hisashi Yamazaki, 
Nobuyoshi Yamada, Kazuhiko Kuroki} 
\address{Department of Applied Physics \& Chemistry, The University of 
Electro-Communications,
Chofu, Tokyo 182-8585, Japan}
\author{Masamitsu Watanabe}
\address{RIKEN/SPring-8,
Kouto 1-1-1, Mikazuki, Sayo, Hyogo 679-5148, Japan}
\author{Tamio Oguchi}
\address{Department of Quantum Matter, ADSM, Hiroshima University,
Higashihiroshima-shi, Hiroshima 739-8530, Japan}
\author{Sergey Lee, Ayako Yamamoto, Setsuko Tajima,}
\address{Superconductivity Research Laboratory, International Superconducting 
Technology Center,
Koto, Tokyo 135-0062, Japan}
\author{Yuji Umeda, Shin Minakawa, Noriaki Kimura, Haruyoshi Aoki,}
\address{Center for Low Temperature Science, Tohoku University,
Sendai, Miyagi 980-8578, Japan}
\author{Shigeki Otani}
\address{Advanced Materials Laboratory, NIMS,
Tsukuba, Ibaraki 305-0044, Japan}
\author{Shik Shin}
\address{The Institute for Solid State Physics (ISSP), The University of Tokyo,
Kashiwa, Chiba 277-8581, Japan\\
RIKEN/SPring-8,
Kouto 1-1-1, Mikazuki, Sayo, Hyogo 679-5148, Japan}
\author{Thomas A. Callcott}
\address{Department of Physics, University of Tennessee,
Knoxville, TN 37996}
\author{David L. Ederer}
\address{Department of Physics, Tulane University,
New Orleans, LA 70118}
\author{Jonathan D. Denlinger}
\address{Advanced Light Source, Lawrence Berkeley National Laboratory,
Berkeley, CA 94720}
\author{Rupert C.C. Perera}
\address{Center for X-ray Optics, Lawrence Berkeley National Laboratory,
Berkeley, CA 94720
}
\date{Accepted to PRB: 2 July 2003}
\abstract{\small
The effect of electron correlation (EC) on the electronic structure in
MgB$_2$, AlB$_2$ and ZrB$_2$,
is studied by examining the partial density of states (PDOS) of
B-2$p\sigma$ and $p\pi$ orbitals using the polarization dependence of x-ray
emission and absorption spectra.
The discrepancies between observed and calculated PDOSs  
cannot be attributed to EC effects.
The present results suggest that the EC effect is less than the
experimental error ($\sim$ 0.2 eV),
which indirectly supports a scenario that electron-phonon interaction
plays an essential role in the occurrence of superconductivity.
\\
\\
74.70.Ad, 74.25.Jb, 78.70.Dm, 78.70.En \\
\hspace{10mm}
}
\maketitle
\narrowtext
Since the discovery of superconductivity in MgB$_2$ with $T_{\rm c}$ of 39
K\cite{Nagamatsu},
a large number of researches from experimental
\cite{Bud'ko,Kotegawa,Tsuda,Bouquet,Szabo,Quilty,Callcott,Nakamura1,Kurmaev,Nakamura2,Schuppler} 
and theoretical
\cite{Kortus,An,Imada,Liu,Shulga,Yamaji,Choi} point of view have 
been performed on MgB$_2$ and on a series of isostructural diborides.
Most of these studies suggest that MgB$_2$ is a phonon-mediated
BCS type superconductor.
Bud'ko {\it et al.} reported a boron isotope effect with
$\alpha$=0.26.\cite{Bud'ko}
The temperature dependence of $^{11}$B-nuclear spin relaxation
rate, $1/T_1$, shows an exponential decay in the superconducting state
revealing a tiny coherence peak just below $T_{\rm c}$, which means
that MgB$_2$ is an $s$-wave superconductor
with a large band gap.\cite{Kotegawa}
On the other hand, high resolution photoemission spectroscopy and
specific heat measurement of MgB$_2$ suggest the two superconducting 
gaps.\cite{Tsuda,Bouquet}
The tunneling experiment also suggest that two gaps of about 2-3 and 7 
meV.\cite{Szabo}
Recent Raman study on the single crystalline MgB$_2$ assigned the two gaps
to a large one (6.5 meV) of the $\sigma$ band and a small one
(1.5 meV) of the $\pi$ bands.\cite{Quilty}
These results are in contradiction with a scenario that MgB$_2$ is a
simple $s$-wave superconductor.
Theoretical band calculations\cite{Kortus,An}, in the early stage,
have suggested that
the dimensionless electron-phonon coupling (EPC) constant $\lambda \sim$0.7,
which
can give a high $T_{\rm c}$ of $\sim$40 K if the Coulomb potential $\mu^*$ is
very small.
On the other hand, two-band mechanisms based on inter-band
electron-correlation (EC)
have been proposed.\cite{Imada,Liu,Shulga,Yamaji}
In these mechanisms, inter-band EC ($\sigma$ and $\pi$ bands considered
by Imada, and bonding and antibonding $\pi$-bands by Yamaji)
enhances $T_{\rm c}$ from a
conventional BCS (EPC based) value, and the mechanisms have relationship
with the
experimental results that suggest the two superconducting
gaps\cite{Tsuda,Bouquet,Szabo,Quilty}.
In order to understand the high $T_{\rm c}$ of MgB$_2$, therefore, it is
necessary
to get information on the EC effects in the superconducting MgB$_2$ and
non-superconducting other diborides.
The density of state gives important information on the EC effects.
The partial density of states (PDOS) of boron have been measured by
x-ray absorption (XAS) and x-ray emission (XES) spectroscopy near B-$K$
edge of polycrystalline MgB$_2$ and other AlB$_2$-type compounds,
in which the observed PDOS agrees well with the band
calculations.\cite{Callcott,Nakamura1,Kurmaev}
Furthermore, angle resolved photoemission spectroscopic
(ARPES)\cite{Uchiyama} and
de Haas van Alphen (dHvA) effect\cite{Yelland} studies were performed
on the single crystalline MgB$_2$ sample\cite{Lee}.
The ARPES spectra along the $\Gamma$-$K$ and $\Gamma$-$M$ directions show
three dispersive curves that can be assigned to
theoretically predicted $\sigma$ and $\pi$ bands.
However, some predicted bands were not observed.
In addition, a small parabolic-like band is observed around the $\Gamma$ point,
which can not be explained by band calculations.
Because this technique is quite surface sensitive,
the results may not represent the bulk-electronic structure.
On the other hand, dHvA technique is useful to probe the bulk-electronic
structure.
Yelland et al. reported that only three dHvA frequencies were resolved
among four Fermi surfaces predicted theoretically.
The derived three dHvA frequencies and the large effective mass are, however,
explained by precise band calculation,\cite{Mazin,Rosner} the calculation 
insists
that the bands near $E_{\rm F}$
should shift with decreasing number of holes near $E_{\rm F}$.
They pointed out that the discrepancies between the experimental results and
the band calculations may be caused by EC effects or beyond-LDA effect.
Furthermore, several authors have proposed a model based on a weak 
electron-phonon coupling\cite{Marsiglio} that is consistent with the
optical conductivity and DC resistivity studies of $c$-axis oriented
MgB$_2$ films\cite{Tu}.
Thus, it is necessary to investigate the significance of the EC effects,
which can play an important role on the appearance of
high $T_{\rm c}$ in MgB$_2$.
XAS and XES measurements of single crystal are quiet useful for this
purpose because these techniques
give PDOS which can reflect the existence of strong EC.
XAS and XES of single crystalline AlB$_2$ and XAS of single crystalline
Mg$_x$Al$_{1-x}$B$_2$ were already performed,\cite{Nakamura2,Schuppler}
in which a good agreement between the observation and the band calculation
was reported.
In this paper, we report a direct observation of PDOS of B-2$p\sigma$ and
2$p\pi$
by polarization dependent XES and XAS near B-$K$ edge using single crystalline
MgB$_2$, AlB$_2$ and ZrB$_2$ samples.
Comparing the observed PDOS with the first principle band calculation
results\cite{Oguchi},
we examine the significance of EC effects in MgB$_2$.

The single crystalline MgB$_2$ samples were grown in BN container under
high pressure.
\cite{Lee}
AlB$_2$ and ZrB$_2$ crystals were prepared by Al-flux\cite{Nakamura2} and
FZ methods\cite{Ohtani}, respectively.
The XES measurement was performed at BL-2C in KEK-PF,
in which the energy of the incident photons is about 400 eV.
The energy resolution of XES spectrometer is about 0.2 eV.
The XAS spectra were measured at BL-8.0.1 of Advanced Light Source (ALS)
in LBNL by the partial fluorescence yield (PFY) method.
The energy resolution of the incident photons is about 0.1 eV.
PDOSs of each B-2$p\sigma$ and 2$p\pi$ orbitals are derived from
polarization dependence of XES and XAS spectra.\cite{Nakamura2}

Figure 1 shows the observed partial density of states (PDOS) of
B-2$p\sigma$ [Fig. 1(b)] and $p\pi$ states [Fig. 1(c)] from
observed polarization-dependent XES and XAS spectra with the results of
band calculation.
Solid and dotted lines in Fig.1 are the results of the first principle
band calculations (FLAPW method) by Oguchi\cite{Oguchi},
which are convoluted by gaussian function with FWHM of the experimental
resolution.
Solid and open circles (or squares) represent occupied and empty states
of $p\sigma$ (or $p\pi$), respectively.
It is clearly seen that the Fermi energy $E_{\rm F}$ [A in Figs. 1(b) and
(c)] measured
from B-1$s$ core level of MgB$_2$ is 186.4 eV, which agrees well
with the previous reports.\cite{Nakamura1}
In the Figs. 1(b) and (c), the theoretical $E_{\rm F}$ value is set to the
experimental
$E_{\rm F}$ value, 186.4 eV.
A sharp peak B in XES spectrum [Fig.1(b)] is observed at around $E-E_{\rm
F}=-2.4$ eV.
Observed PDOS of $p\sigma$ in XES spectrum steeply decreases at $E_{\rm F}$ and
a considerable amount of PDOS just above $E_{\rm F}$ is observed in XAS
spectrum.
The $p\sigma$-PDOS near $E_{\rm F}$ disappears above 0.6 eV (C),
and there is almost no $p\sigma$-PDOS in the energy region D
(0.6 eV$<E - E_{\rm F}<$3.6 eV).
Figure 1(c) shows XES $(\rule[0.2mm]{2mm}{2mm})$ and XAS $(\Box)$ of
B-2$p\pi$ of MgB$_2$.
Observed PDOS of $p\pi$ shows a broad metallic state except a large sharp
peak G
at 5.6 eV.
The overall features of observed XAS near $E_{\rm F}$ and XES are well
reproduced
by the band calculation.
However, in whole energy region, some discrepancies are observed as follows.
Observed peak B is lower than the theoretical prediction by 0.3 eV.
The value of observed pseudo-gap is about 3 eV in contrast to the prediction
of about 4 eV.
Peaks, E$_1$ and E$_2$ in $p\sigma$-XAS and G in $p\pi$-XAS, are not reproduced
by the band calculation.
Before going into detailed comparison between the theory and the experiment,
let us show the results AlB$_2$.

Figure 2 shows PDOS of AlB$_2$ with the same symbols of MgB$_2$ in Fig.1.
The observed $E_{\rm F}$ is estimated to be 187.5 eV which agrees well with the
previous report.\cite{Nakamura1,Nakamura2}
The value of $E_{\rm F}$ is slightly lower than the theoretical prediction
by 0.6 eV,
but the small shift is explained by the lack of Al atoms by 0.07 from the
stoichiometric AlB$_2$.\cite{Nakamura2}
As in MgB$_2$, overall shapes of experimental $p\sigma$- and $p\pi$-PDOS
are in good agreement with the band calculation results.
Especially, in AlB$_2$, it is found that a detailed shape of PDOS including
a pseudo-gap in the empty state is in good agreement with the theoretical
prediction within the experimental resolution.
This is in contrast with the case of MgB$_2$.

Figure 3 shows $p\sigma(\circ)$- and $p\pi(\bullet)$-PDOS of ZrB$_2$ derived
from XES spectra.\cite{Watanabe}
In sharp contrast to the PDOSs of MgB$_2$ and AlB$_2$, the $p\pi$-PDOS of
ZrB$_2$ shows clearly-resolved two-large peaks at about 184.3 eV and 185.4 eV,
respectively.
The $p\sigma$-PDOS also shows two peaks at about 183.0 eV and 184.3 eV.
Both PDOSs decrease with increasing energy, but the small Fermi edge is
observed in both PDOSs.
The $E_{\rm F}$ is estimated to be 188.1 eV.
The solid and dotted lines are the theoretical ones with the experimental
$E_{\rm F}$ value.
As in MgB$_2$ and AlB$_2$, even though the PDOS shapes are different,
the observed PDOSs of ZrB$_2$ are well reproduced by the first principle
band calculation.
The detailed comparison between observed PDOS and theoretical ones is as
follows.

As mentioned before, overall shapes of the
observed PDOSs of these compounds are
roughly reproduced by the band calculations, but some discrepancies are
pointed out in MgB$_2$ and AlB$_2$.
A sharp peak B in MgB$_2$, which is due to van Hove singularity (VHS) of
$p\sigma$ band
at $M$- and $L$-points, slightly shifts from the theoretical prediction by
about
$-0.3$ eV.
An energy, measured from $E_{\rm F}$, of bonding $p\sigma$-top at
$\Gamma$-point
[C in Fig. 1(b)] is about 0.6 eV in MgB$_2$ and $-1.0$ eV in AlB$_2$,
respectively.
It agrees with the theoretical prediction in MgB$_2$, and agrees with the
prediction
in AlB$_2$ assuming the $E_{\rm F}$ shifting.\cite{Nakamura2}
However, in MgB$_2$, observed anti-bonding $p\sigma^*$ PDOSs E$_1$ and E$_2$
are higher than the theoretical ones F$_1$ and F$_2$.
This means the observed pseudo-gap located at region D is smaller than
the theoretical prediction in MgB$_2$ by about 1 eV.
On the other hand, in AlB$_2$, one can see an excellent agreement between
observed and theoretical PDOS around the pseudo-gap of about 3 eV.
The values of observed pseudo-gap of both compounds are the same
(3 eV).
In AlB$_2$, there is no characteristic structure in PDOS above $E_{\rm F}$+
5 eV.
Therefore, it seems that there is no discrepancy between experimental and
theoretical PDOS in AlB$_2$ compound even in the high energy region.

The large sharp peak G is due to $p\pi^*$ resonant state of the sample surface
or of some oxides of the surface.\cite{Callcott}
And it does not appear in a polished-large single-crystal of AlB$_2$.
For AlB$_2$ single crystal, in order to remove the Al-flux on the surface,
the crystals were polished.\cite{Nakamura2}
Therefore the fluorescence spectrum will be free from the surface oxidation.
But the size of MgB$_2$ single crystals is too small to remove surface oxides
by polishing.
Then the small amount of oxides leads to the resonant peak G in MgB$_2$ spectrum.
The present observation of PDOS of $p\pi$ band also agrees with the theoretical one
except for the surface states mentioned above.
The present results indicate that the experimental PDOSs are reproduced
by the band calculation in the energy region of $E < E_{\rm F}$+5 eV in
both diborides.

One might consider that the discrepancy between experiment and 
the theory in the XAS regime of MgB$_2$ may be 
due to EC effects that is not properly taken 
into account in LDA band calculations.
However, the EC effects generally tend to {\it widen} the gap,
while in the present case, the experimental band gap is {\it narrower} than
the theoretical one. Then, this discrepancy between the experiment and the 
theory may be attributed to the fact that 
the band calculation deals with the {\it ground state} of the system.
A possible reason for the gap narrowing 
might be due to an excitonic effect\cite{comment} that arises in the excited 
states of the XAS process, which is not taken into account in the 
band calculation.

%
%
In a previous paper\cite{Nakamura1},
we insisted that a rigid band picture is valid for the relation between
MgB$_2$ and AlB$_2$.
The present detailed PDOSs of both compounds do not deny the rigid band
picture,
but suggest a small discrepancy between both compounds, i.e., anti-bonding
$p\sigma^*$ states is lower than the theoretical prediction in MgB$_2$ but
that in AlB$_2$
is in agreement with the theoretical one.
In ZrB$_2$, the observed $p\pi$-PDOS structure is similar to the
theoretical PDOS of
Zr-4$d$.\cite{Watanabe}
The high energy peak at 184.3 eV of $p\sigma$-PDOS is also similar to the
Zr-4$d_{yz,zx}$ PDOS, but the low energy PDOS at 183.0 eV is considered to
be based
on the covalent character of B-B bonding in basal plane.
As mentioned in the introductory part, there are two types of theoretical
two-bands
model based on electron-phonon\cite{An} and inter-band EC
mechanisms\cite{Imada,Yamaji}
in MgB$_2$ superconductivity.
The present result indicates that the EC is smaller than the value of the
present energy resolution ($\sim$0.2 eV) in MgB$_2$, AlB$_2$ and ZrB$_2$
compounds.
The inter-band EC has a possibility to enhance the
phonon-mediated $T_{\rm c}$.\cite{Imada,Yamaji}
If a small inter-band EC effect that can not be detected in
our experiment enhances the high $T_{\rm c}$,
the present result does not contradict with these propositions.

To summarize, in order to examine the electron correlation (EC) effect
in the diborides, we have performed direct measurement of PDOS of
B-$2p$ in single-crystalline MgB$_2$, AlB$_2$ and ZrB$_2$
using polarization-dependent XES and XAS measurements.
Although there are some discrepancies between observed PDOSs and
theoretical ones, the first principle band calculation reproduces
well the overall features of observed $p\sigma$- and $p\pi$-PDOSs.
In superconducting MgB$_2$, a considerable amounts of $p\sigma$-hole state
near the Fermi energy is clearly observed.
The pseudo-gap of $p\sigma$ band is observed in MgB$_2$ and AlB$_2$ compounds
in sharp contrast to the broad metallic state of the B-2$p\pi$ bands.
The observed gap values of about 3 eV are same in both compounds,
which is smaller than the theoretically predicted value for MgB$_2$ and
is consistent with it for AlB$_2$.
Because the band calculation describes the ground state,
it may be plausible that the calculation reproduces the experimental PDOS
only for $E < E_{\rm F}$+5 eV in both compounds.
In ZrB$_2$, the observed PDOSs reproduced well by the calculation,
suggest strong hybridization between B-2$p$ and Zr-4$d$ orbitals.
The observed discrepancies are contrary to the EC effects.
The present results suggest that the EC effect is less than the
experimental error ($\sim$ 0.2 eV),
which indirectly supports a scenario that electron-phonon interaction
plays an essential role in the occurrence of superconductivity in MgB$_2$.
However, a possibility of a small inter-band electron-correlation effect
which supports
the phonon mediated superconductivity, still remains.

J.N. and S.N. thank to Dr. Y. Kobayashi of RIKEN for the support
with the synthesis of AlB$_2$ single crystals.
K.K. acknowledges Dr. J. Yamauchi of Toshiba Corporate R\&D Center 
for valuable discussions.
This work was supported by Office of Basic Energy Science of the
Department of Energy under contract Nos. DE-AC03-76SF00098 and W-7405-Eng-48,
National Science Foundation (NSF) Grant No. DMR-9017996 to the University of
Tennessee and a Grant-in-Aid No. 14540330 and No.13CE2002 (for COE Research)
of the Ministry of Education, Culture, Sports, Science and Technology of
Japan.
This work was performed under the approval of the KEK-PF Advisory Committee
(Proposal No. 2001U004).
D.L.E. and T.A.C. also wish to acknowledge support from NSF Grant DMR-9801804.

\begin{figure}
\caption{
The partial density of states (PDOS) of $p\sigma$ and $p\pi$ of MgB$_2$.
(a) The theoretical PDOS derived from FLAPW method broadened with
experimental resolution.
The solid and dotted lines are PDOS's of $2\times p\sigma$ and $p\pi$, 
respectively.
(b) The experimental PDOS of $2\times p\sigma$, occupied one (solid circle) and
empty one (open circle).
(c) The experimental PDOS of $p\pi$, occupied one (solid square) and empty
one (open square).
}\label{FIG1}
\end{figure}
\begin{figure}
\caption{
The partial density of states (PDOS) of $p\sigma$ and $p\pi$ of AlB$_2$.
(a) The theoretical PDOS derived from FLAPW method broadened with
experimental resolution.
The solid and dotted lines are PDOS's of $2\times p\sigma$ and $p\pi$, 
respectively.
(b) The experimental PDOS of $2\times p\sigma$, occupied one (solid circle) and
empty one (open circle).
(c) The experimental PDOS of $p\pi$, occupied one (solid square) and empty
one (open square).
}\label{FIG2}
\end{figure}
\begin{figure}
\caption{
The partial density of states (PDOS) of $2\times p\sigma (\circ)$ and $p\pi
(\bullet)$ of ZrB$_2$.
The solid and dotted lines are theoretical PDOSs of $2\times p\sigma$ and 
$p\pi$,
respectively,
which are broadened with experimental resolution.
}\label{FIG3}
\end{figure}


\begin{thebibliography}{99}
\bibitem{comment} Two points should be mentioned here. First, although 
the excitonic effect is a Coulombic effect, it is not essentially an EC effect.
Namely, the excitonic shift of the energy can essentially be evaluated 
within a mean field scheme if we take into account the 
excited states of the system. Secondly, one may consider that 
excitonic effects may not be so large in metallic systems, 
where the screening of the Coulomb potential is prominent. 
If that is indeed the case, a clearcut reason for 
the discrepancy between theory and the experiment remains to be 
an open question. However, since EC effects generally tend to widen the gap,
we believe our scenario that the EC effects is not so large in MgB$_2$ 
remains unchanged in any case.
\end{thebibliography}
\end{document}